\documentstyle[pra,aps,graphicx,twocolumn]{revtex}
\begin{document}
\title{Magneto-optical control of bright atomic solitons}
\author{S.\ P\"otting, O. Zobay, P. Meystre, and E.\ M.\ Wright}
\address{Optical Sciences Center, University of Arizona, Tucson,
Arizona 85721}
\address{email: sierk.poetting@wotan.opt-sci.arizona.edu} 
\address{phone: +1 (520) 621--2077,
         fax: +1 (520) 621-4358}

\maketitle
\begin{abstract}
In previous work we showed that bright atomic solitons can arise
in spinor Bose--Einstein condensates in the form of gap solitons
even for repulsive many--body interactions.  Here we further
explore the properties of atomic gap solitons and show that their
internal structure can be used to both excite them and control
their center--of--mass motion using applied laser and magnetic
fields.  As an illustration we demonstrate a nonlinear
atom--optical Mach-Zehnder interferometer based on gap solitons.
\end{abstract}
\narrowtext
\section{Introduction}
\label{introduction}
The experimental demonstration of Bose-Einstein condensation in
atomic vapors has rapidly lead to spectacular new advances in atom
optics. In particular, it has enabled its extension from the
linear to the nonlinear regime \cite{Lenz93}, very much like the laser led to
the development on nonlinear optics in the 1960s. It is now well
established that two-body collisions play for matter waves a role
analogous to that of a Kerr nonlinear crystal in optics. The first
experimental verification of this analogy was four-wave
mixing \cite{Deng99}. In addition, it is also possible to
nonlinearly mix optical and matter-waves, as demonstrated recently
in matter-wave superradiance \cite{Inou99,Moor99a} and in the first realization of a
phase-coherent matter-wave amplifier \cite{Kozu99}. In addition, it is known
that the nonlinear Schr\"odinger equation which describes the
condensate in the Hartree approximation support soliton solutions \cite{Scot73}.
For the case of repulsive interactions normally encountered in BEC
experiments, the simplest solutions are dark solitons, that is,
``dips'' in the density profile of the condensate 
\cite{Rein97,Scot98,Dum98,Dobr99}. These dark
solitons have been recently demonstrated in two experiments
\cite{Dens00,Burg99} which appear to be in good agreement with the
predictions of the Gross-Pitaevskii equation.

While very interesting from a fundamental physics point-of-view,
dark solitons would appear to be of limited interest for
applications such as atom interferometry, since what would be
desirable is to achieve the dispersionless transport of a
spatially localized ensemble of atoms, rather than a "hole." In
that case, bright solitons are much more interesting. However, the
problem is that large condensates are necessarily associated with
repulsive interactions, for which bright solitons might appear to
be impossible since the nonlinearity cannot compensate for the
kinetic energy part (diffraction) in the atomic dynamics. While
this is true for atoms in free-space, this is not the case for
atoms in suitable potentials, eg. optical lattices. This is
because in that case, it is possible to tailor the dispersion
relation of the atoms in such a way that their effective mass
becomes negative. For such negative masses, a repulsive
interaction is precisely what is required to achieve soliton
solutions. This result is known from nonlinear optics, where such
soliton solutions, called gap solitons, have been predicted and
demonstrated \cite{Chri89,Ster94}.

We showed in previous work \cite{Zoba99} that such bright solitons are also
possible in matter-wave optics using spinor condensates, but did
not discuss explicitly how to excite and control them. The present
paper addresses these questions, and shows that a combination of
optical and magnetic fields can be used to generate solitons of
various velocities, and subsequently, to control them, split and
recombine them, etc. This magneto-optical control results from the
use of spinor condensates, and can be achieved at minimal cost in
terms of atomic loss. In particular, we illustrate how to realize
a Mach-Zehnder interferometer for bright atomic solitons. This
opens up the way to intriguing new ways to manipulate and
transport coherent matter in ways which complement those offered
by standard optical tweezers or by atomic wave guides 
\cite{Key00,Dekk00}.

Amongst Dan Walls many contributions to nonlinear and atom optics
he was one of the first to recognize the emerging area of
nonlinear atom optics, and he introduced the idea of atomic
solitons traveling in laser beams already in $1994$ along with
Zhang and Sanders \cite{Zhan94}.  The current wave of interest in atomic
solitons therefore finds its origin in Dan's seminal work, and we
all greatly miss the originality, insight, and energy he brought
to the field.

The paper is organized as follows: Section \ref{model} briefly
reviews the analysis leading to the predictions of gap solitons
\cite{Zoba99}, while section \ref{svea} discusses some of their
most important characteristics. Section \ref{control} exploits
these properties to develop tools to excite gap solitons and
control their dynamical behavior, leading to the demonstration of
an atomic Mach--Zehnder interferometer in Section
\ref{machzehnder}. Finally, section \ref{conclusion} is a summary
and outlook. While the main text uses a simple model of
light-matter interaction for notational clarity, a more realistic
coupling scheme, using the full hyperfine structure of the Sodium
$3S_{1/2}$--$3P_{3/2}$ transition, is presented in Appendix
\ref{appendix}.
\section{Physical Model}
\label{model}
\begin{figure}
\includegraphics*[width=0.95\columnwidth]{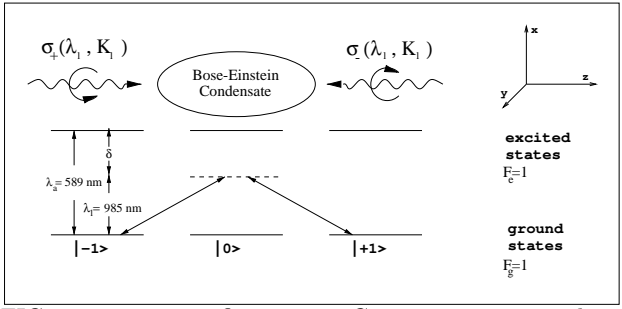}
\caption{$\sigma_+$--$\sigma_-$ configuration: Counterpropagating
  laser fields induce far--off resonance Raman coupling between
  hyperfine states.}
\label{figgeometry}
\end{figure}

To set the stage for our analysis, we first briefly review the
main ingredients of the theory of gap solitons of Ref.
\cite{Zoba99}. The system we consider consists of a Bose-Einstein
condensate interacting with two counterpropagating, focused
Gaussian laser beams of equal frequency $\omega_l$ but opposite
circular polarizations, see Fig. \ref{figgeometry}. The optical
dipole potential associated with the applied laser beams is
assumed to provide tight transverse confinement for the BEC in the
$(X,Y)$ plane, thereby forming a cigar shaped condensate of
transverse cross-sectional area $A_T$. In the following, we
confine our discussion to the one-dimensional dynamics of the BEC
along the $Z$-axis for simplicity.

In addition to supplying a transverse optical potential, the laser
beams can drive two-photon transitions between different Zeeman
sublevels of the atomic ground state. For illustrative purposes we
consider the case of Sodium and the two-photon coupling of the
Zeeman sublevels $|-1\rangle=| F_g = 1, M_g=-1\rangle$ and
$|1\rangle=| F_g = 1, M_g=1\rangle$. For example, starting in the
$|-1\rangle$ state this process involves the absorption of a
$\sigma_+$ photon from the right propagating laser beam followed
by emission of a $\sigma_-$ photon into the left propagating laser
beam.

We must of course assume that the excited states involved in the
atom-field interaction are far-detuned from the applied laser
frequency, a necessary requirement to avoid the detrimental
effects of spontaneous emission. This however raises a serious
issue, since for the alkali atoms that we have in mind the
two-photon coupling strength vanishes in the limit when the
detuning is large compared to the excited--state hyperfine
splitting \cite{Deut98,Hagl99a}. But this difficulty can be
circumvented: We show in the Appendix that by using a four-photon
scheme involving an additional $\pi$-polarized field incident
perpendicular to the $Z$-axis, one can achieve an effective
two-photon coupling between the states $|\pm 1\rangle$ which
survives in the limit of large detunings. We therefore proceed
with our two-photon coupling model and refer the interested reader
to the Appendix for details.

By restricting our attention to the coupled states $|\pm 1\rangle$
the effective single--particle Hamiltonian for our model system
can be written as \cite{Zoba99,Cohe92}
\begin{equation}
\label{eq:heff} H_{\mathit{eff}} = \frac{P_Z^2}{2m}+g\hbar\delta'
                 \left[
                    \left|1\right\rangle\left\langle -1\right| e^{2iK_l Z} +
                    \left|-1\right\rangle\left\langle 1\right| e^{-2iK_l Z}
                 \right ],
\end{equation}
where we have omitted constant light-shift terms.  Here $P_Z$ is
the atomic center--of--mass momentum operator along $Z$, $m$ the
mass of the atom, $K_l=\omega_l/c$ is the magnitude of the field
wave vector along $Z$, $g$ is a coupling constant between the
ground and excited states characteristic of the atom and
transition involved, $\delta' =
{\mathcal{D}}^2{\mathcal{E}}^2/\hbar^2 \delta$, with the detuning
$\delta = \omega_l - \omega_a$, $\mathcal{E}$ is the laser field
amplitude, and $\mathcal{D}$ is the reduced dipole moment for the
$3S_{1/2}$--$3P_{3/2}$ transition.

The first term of the effective Hamiltonian (\ref{eq:heff})
describes the quantized atomic center-of-mass motion, and the
remaining terms give the effective coupling of the two Zeeman
sublevels via the applied laser fields. The exponential terms
$\exp(\pm 2iK_lZ)$ arise from the fact that the two-photon
transitions involve the absorption of a photon from one light
field and reemission into the other. Finally, introducing a spinor
macroscopic condensate wave function $\Psi(Z,t)
=[\Psi_{1}(Z,t),\Psi_{-1}(Z,t)]^T$ normalized to the number of
atoms $N$, and including the many--body effects via a mean-field
nonlinearity, we obtain the coupled Gross--Pitaevskii equations
\begin{equation}
\label{gpcop} i\hbar\frac{\partial\Psi}{\partial t}
=H_{\mathit{eff}}\Psi + U|\Psi|^2\Psi ,
\end{equation}
where $U=4\pi\hbar^2a_{sc}/m$, $a_{sc}$ is the s-wave scattering
length, $|\Psi|^2=|\Psi_1|^2+|\Psi_{-1}|^2$, and we have assumed
that the magnitude of the self- and cross-nonlinearities are equal
for simplicity.

It is convenient to re-express Eq.\ (\ref{gpcop}) in dimensionless
form by introducing the scaled variables $\tau=t/t_{c}$, $z
=Z/l_{c}$ and $\psi_j=\Psi_j/\sqrt{\rho_{c}}$ where
\begin{equation}
  \label{eq:scalefactors}
  t_c = \frac{1}{g \delta'}, \quad
  l_c = \frac{t_c \hbar K_l}{m}, \quad
  \rho_c = \left|\frac{g\hbar\delta'}{U}\right|.
\end{equation}
Equations (\ref{gpcop}) then become
\begin{eqnarray}
  \label{eq:scaled}
  i \frac{\partial}{\partial \tau}
  \left(\begin{array}{c} \psi_1 \\ \psi_{-1} \end{array}\right)
  & = &
  \left(\begin{array}{cc}
        \displaystyle{-M\nabla^2} &
        \displaystyle{e^{2ik_lz}} \\
        \displaystyle{e^{-2ik_lz}} &
        \displaystyle{-M\nabla^2}
  \end{array}\right)
  \left(\begin{array}{c} \psi_1 \\ \psi_{-1} \end{array}\right) \nonumber\\
  & + & \mbox{sgn}\left(g\delta'/U\right)\left|\psi\right|^2
  \left(\begin{array}{c} \psi_1 \\ \psi_{-1} \end{array}\right),
\end{eqnarray}
where $M=g \delta' m/2 \hbar K_l^2$ is a mass-related parameter
such that $k_l = K_l l_c = 1/2M$.  Throughout this paper we use a
characteristic laser intensity of $I = 50 \mbox{ W/cm}^2$ for each
of the two counter-propagating laser beams and a wavelength of
$\lambda_l = 985 \mbox{ nm}$. In addition we use $g=-1/4$, which
is simply the Clebsch-Gordan coefficient for the $F_g=1
\leftrightarrow F_e=1$ transition.  This results in the
characteristic scale values $t_c=685 \mbox{ }\mu\mbox{s},\quad
l_c=12.1 \mbox{ }\mu\mbox{m},\quad \rho_c=8.59 \mbox{
}\mu\mbox{m}^{-3}$ and $M=0.0065$.
\section{Gap Solitons}
\label{svea} The spatially modulated coupling between the optical
fields and the condensate induces a single-particle band structure
with regions of negative effective mass. As mentioned in the
introduction, this leads to the possibility of bright atomic
solitons even for repulsive interactions \cite{Zoba99}. Their
energy lies in forbidden gaps of the linear band structure, hence
the name gap solitons.
\subsection{Analytical Solutions}
\label{analyticalsolutionsvea}
Approximate analytic expressions for the gap solitons can be
obtained by expressing the spinor condensate components as
\begin{equation}
  \label{eq:factoroutsvea}
  \psi_{\pm1}(z,\tau)
  =
  e^{\pm ik_lz}e^{-i\tau/4M}\phi_{\pm1}(z,\tau),
\end{equation}
where the field envelopes $\phi_{\pm1}$ are assumed to be slowly
varying in space compared to $1/k_l$.  Neglecting then the
second-order spatial derivatives yields the coupled partial
differential equations
\begin{eqnarray}\label{gapp}
i\left (\frac{\partial}{\partial\tau}\pm \frac{\partial}{\partial
z} \right ) &&\left(\begin{array}{c} \phi_1
\\ \phi_{-1}\end{array} \right)=\left(\begin{array}{cc} 0 & 1\\ 1
& 0 \end{array}\right) \left(\begin{array}{c} \phi_1 \\ \phi_{-1}
\end{array} \right) \nonumber \\ &&\pm (|\phi_1|^2+|\phi_{-1}|^2)
\left(\begin{array}{c} \phi_1 \\ \phi_{-1}
\end{array} \right)  ,
\end{eqnarray}
where the choice $\pm 1=\mbox{sgn}(g\delta^\prime/U)$. For a
red-detuned laser and our choice of $g$ this becomes $\pm
1=\mbox{sgn}(U)$. Aceves and Wabnitz \cite{Acev89} have shown that
these dimensionless equations have the explicit two-parameter gap
soliton solutions (see also Ref. \cite{Ster94})
\begin{eqnarray}
\phi _{1}&=&\pm \frac{\sin(\eta)}{\beta\gamma\sqrt{2}}
            \left( -\frac{e^{2\theta }+e^{\mp i\eta }}
             {e^{2\theta }+e^{\pm i\eta }}\right) ^{v}sech
            \left( \theta \mp \frac{i\eta }{2}\right)
            e^{\pm i\sigma }  ,
\nonumber \\ \phi_{-1}&=&-\frac{\beta\sin(\eta)}{\gamma\sqrt{2}}
             \left( -\frac{e^{2\theta }+e^{\mp i\eta }}
              {e^{2\theta }+e^{\pm i\eta }}\right) ^{v}sech
             \left( \theta \pm \frac{i\eta }{2}\right)
             e^{\pm i\sigma }  ,
\label{GapEnv}
\end{eqnarray}
with $-1<v<1$ is a parameter which controls the soliton velocity,
$0<\eta<\pi$ is a shape parameter, and
\begin{equation}
\label{eq:param1} \beta =\left( \frac{1-v}{1+v}\right)
^{\frac{1}{4}}  , \quad \gamma =\frac{1}{\sqrt{1-v^{2}}}  ,
\end{equation}
\begin{equation}
\label{eq:param2} \theta =-\gamma\sin(\eta)(z-v\tau )  , \quad
\sigma =-\gamma\cos(\eta)(vz-\tau ).
\end{equation}
Since we are interested in creating bright solitons in the
presence of repulsive interactions we restrict ourselves to
$\mbox{sgn}(U)=+1$, corresponding to the choice of the upper sign
in the analytic solutions.

The characteristic length scale associated with the solitons is
$l_c$, so that the approximate solitons (\ref{GapEnv}) are valid
for $K_ll_c = 1/2M >> 1$. This inequality is well satisfied for
our choice of parameters, which gives $1/2M=76.9$.
\subsection{Characteristic Properties}
\label{characteristics}
\begin{figure}
\includegraphics*[width=0.95\columnwidth]{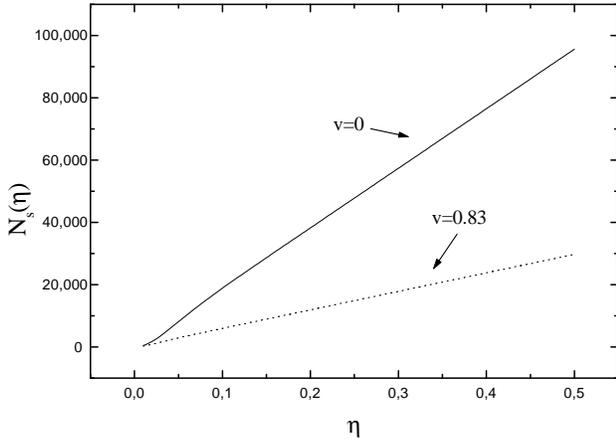}
\caption{Total Number of atoms as a function of parameter $\eta$ for a
         stationary soliton and a
         moving soliton with $v=0.83$ (i.e
         $1.47\mbox{ cm/s}$). Moving solitons always
         contain less atoms than the stationary solitons.
         The transverse width was chosen to be $24.2\mbox{
           }\mu\mbox{m}$, corresponding to an effective transverse
         area $A_T=919$ $\mu\mbox{m}^{2}$.}
\label{figneta}
\end{figure}

\begin{figure}
\includegraphics*[width=0.95\columnwidth]{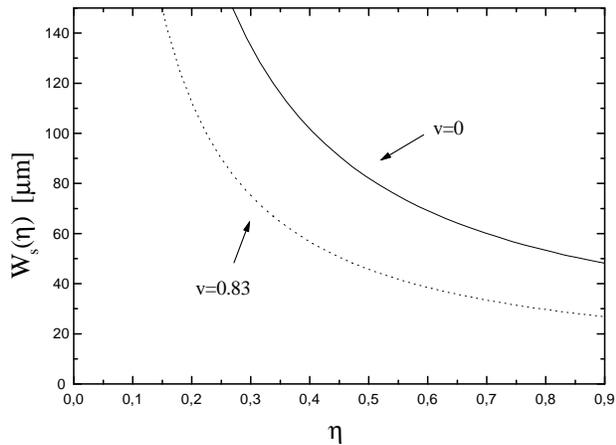}
\caption{Width of the soliton as a function of parameter
         $\eta$ for a stationary soliton and a moving soliton with
         $v=0.83$ (i.e. $1.47\mbox{ cm/s}$). For the regime of small $\eta$
         we obtain widths of 40--100 $\mu\mbox{m}$.}
\label{figwidtheta}
\end{figure}

From the dependence of the hyperbolic--secant on
$\theta=-\gamma\sin(\eta)(z-v\tau)$ in Eqs. (\ref{GapEnv}), we
identify the gap soliton parameter $v=V_g/V_R$ as the group
velocity $V_g$ of the soliton in units of the recoil velocity
$V_R=l_c/t_c=\hbar K_l/m $. Since $-1<v<1$, the magnitude of the
group velocity is bounded by the recoil velocity, which is
$V_R=1.77 \mbox{ cm/s}$ for the present case of Sodium.  From Eqs.
(\ref{GapEnv}), one can extract further important soliton
properties, such as the number $N_s$ of atoms in the soliton shown
in Fig. \ref{figneta}, and the soliton width $W_s=w_sl_c$ shown in 
Fig. \ref{figwidtheta}.
Specifically, the number of atoms in a particular gap soliton is
given by
\begin{equation}
N_s=A_T\int dZ [|\Psi_1(Z,t)|^2 + |\Psi_{-1}(Z,t)|^2], \label{Ns}
\end{equation}
where $A_T$ is the effective transverse area.
Fig. \ref{figneta} illustrates two general soliton properties, namely that
increasing the shape parameter $\eta$ increases the atom number,
but that faster solitons have lower atom number. Similarly, 
Fig. \ref{figwidtheta}
shows that increasing the shape parameter $\eta$ decreases the
soliton width and that faster solitons have narrower widths.

Given that the analytic gap soliton solutions (\ref{GapEnv}) hold
for broad envelopes, we confine our attention to the case $\eta <
1$, with atom numbers in the range $N_s\simeq 10^4-10^5$, and
soliton widths $W_s\simeq 40-100$ $\mu$m for the parameters at
hand.

We can gain further insight into the structure of the gap solitons
by taking the more extreme limit $\eta<<1$ of Eqs. (\ref{GapEnv}).
Using the definitions (\ref{eq:factoroutsvea}), and returning to
dimensional units we have then
\begin{eqnarray}
\Psi_{1}(Z,0)&=&
            \frac{\eta}{\beta\gamma}\sqrt{\frac{\rho_c}{2}}
            sech(Z/W_0)(-1)^v e^{i(K+K_l)Z}  ,
\nonumber \\ \Psi_{-1}(Z,0)&=&
\frac{\beta\eta}{\gamma}\sqrt{\frac{\rho_c}{2}}
             sech(Z/W_0)(-1)^v e^{i[(K-K_l)Z + \pi]}  ,
\label{GapTot}
\end{eqnarray}
where
\begin{equation}
W_s = 3.44W_0=3.44\left (\frac{l_c\sqrt{1-v^2}}{\eta}\right )  ,
\quad K=-\frac{\gamma v}{l_c} . \label{SolPar}
\end{equation}
Here $W_s=3.44W_0$ is the soliton width, the factor $3.44$ being
the numerical conversion from the width of the hyperbolic secant
to the $1/e^2$ width of the distribution, and $K=k/l_c$ a
velocity-dependent wave  vector shift.  This expression agrees
well with the features displayed in Fig. \ref{figwidtheta}, in that the width
decreases with increasing $\eta$ and $v$. The soliton atom number
obtained by combining Eqs. (\ref{Ns}) and (\ref{GapTot}),
\begin{equation}
N_s = 2(A_Tl_c\rho_c)\eta\cdot(1-v^2)  ,
\end{equation}
correctly predicts the scaling properties of Fig. \ref{figneta}.

An essential point to keep in mind is that the gap solitons are
coherent superpositions of the two Zeeman sublevels, and the
approximate solutions (\ref{GapTot}) contain important information
on the phase and amplitude relations that need to be created
between them to successfully excite and manipulate gap solitons.
In particular, they show that there is always a spatially
homogeneous $\pi$ phase difference between the two states.  In
addition, the two components have the spatial wave vectors
\begin{equation}
K_{\pm 1} = K \pm K_l  , \label{Kpm}
\end{equation}
with $K=-\gamma v/l_c$.  For our parameters, $K_l=6.38$
$\mu$m$^{-1}$ and $|K| < 1/l_c = 0.08$ $\mu$m$^{-1}$. Despite the
fact that it is so small, $|K|$ is an important factor since it
controls the soliton velocity. Finally, it follows from dividing
the amplitudes of the two components that
\begin{equation}
\left |\frac{\Psi_1(Z,t)}{\Psi_{-1}(Z,t)} \right |^2 =
\frac{1}{\beta^4} = \left (\frac{1+v}{1-v} \right )  , \label{Rpm}
\end{equation}
which shows that their relative occupation depends on the soliton
velocity parameter $v$. For $v=0$ the sublevels are equally
populated, but as $v\rightarrow 1$ the $|1>$ sublevel has a larger
population, and vice versa for $v\rightarrow -1$.

The characteristic time scale for the evolution of the gap
solitons can be determined from the plane-wave exponential factors
in Eqs. (\ref{GapEnv}). Converting to dimensional form the soliton
period $t_s$ is defined as the time to accumulate a $2\pi$ phase,
or in the limit $\eta\rightarrow 0$
\begin{equation}
t_s = 2\pi t_c\sqrt{1-v^2}  .
\label{solper}
\end{equation}
Physically, $t_s$ corresponds to the internal time scale for the
gap soliton. In order to observe a soliton-like behavior, it is
therefore necessary to investigate the atomic propagation over
several periods.

We conclude this section by noting that Aceves and Wabnitz
\cite{Acev89} have shown that the gap soliton solutions
(\ref{GapEnv}) are stable solutions of Eqs. (\ref{gapp}) in that
they remain intact during propagation, even when perturbed away
from the exact solutions. However, one should remember Eqs.
(\ref{gapp}) are only an approximation to the exact system of Eqs.
(\ref{eq:scaled}) so that in general the gap solitons are solitary
wave solutions only. As such, they are not guaranteed to be
absolutely stable.
\section{Gap soliton control}
\label{control}
\subsection{State manipulation}
\label{tools}
Summarizing the previous section, gap solitons require the right
population in each Zeeman sublevel, a phase difference of $\pi$
between these sublevels, and appropriate plane-wave factors
$e^{iK_{\pm1} Z}$. The shapes of the Hartree wave functions
corresponding to the two Zeeman sublevels are hyperbolic secant,
which we approximate by a Gaussian in the following. They could
for example be initialized in an optical dipole trap
\cite{Stam98}. Manipulating the gap solitons is therefore reduced
to the problem of controlling the populations and phases
throughout the spinor condensate. This is achieved via a
magneto-optical control scheme involving a combination of pulsed
coherent optical coupling and of phase-imprinting using spatially
inhomogeneous magnetic fields.

The coherent optical coupling can be achieved e.g. by a laser
pulse of frequency $\omega_l$ propagating perpendicularly to the
$Z$-axis and with linear polarization perpendicular to that axis.
For sufficiently short pulses, one can neglect changes in the
center-of-mass motion of the atoms during its duration, leading to
a very simple description. We assume for simplicity a plane-wave
rectangular pulse of duration $t_p$ and of spatial extent large
compared to the soliton. The Hamiltonian describing the coupling
between this pulse and the condensate is then the same as in Eq.
(\ref{eq:heff}) without the linear momentum exchange terms
$\exp(\pm 2iK_l Z)$ and the kinetic energy term, and with
$\delta^\prime \rightarrow\delta_p^\prime$. The state of the
system after the pulse is then easily found to be
\begin{eqnarray}
\label{eq:pulsesolution} \left(\begin{array}{c} \Psi_1(t_p) \\
\Psi_{-1}(t_p)\end{array}\right) & = & \left(\begin{array}{cc}
          \displaystyle{\cos\chi} &
          \displaystyle{i\sin\chi} \\
          \displaystyle{-i\sin\chi} &
          \displaystyle{\cos\chi}
\end{array}\right)
\left(\begin{array}{c} \Psi_1(0) \\ \Psi_{-1}(0)\end{array}\right)
\nonumber \\ & \equiv & M_L(\chi) \left(\begin{array}{c} \Psi_1(0)
\\ \Psi_{-1}(0)\end{array}\right).
\end{eqnarray}
where $\chi=g\delta_p^\prime t_p$ is the excitation pulse area and
the operator $M_L$ can be used to control the population transfer
by an appropriate choice of $\chi$.

The required phase relation between the two states can be achieved
via Zeeman splitting. Considering for concreteness a spatially
inhomogeneous rectangular magnetic field pulse of duration $t_B$
we have, neglecting again all other effects,
\begin{equation}
\label{eq:bfieldequation}
i\hbar\frac{\partial}{\partial t}
\Psi_{\pm1}(Z,t)
=
\pm \mu_B g_F \left(B_0+B'Z\right)\Psi_{\pm1}(Z,t),
\end{equation}
where $g_F$ is the Land\'e g--factor of hyperfine ground state,
$\mu_B$ is the Bohr magneton, $B_0$ the spatially homogeneous
component of the magnetic field, and $B'$ its gradient, the
direction of the magnetic field being along the $Z$-axis. The
application of this field results in the state
\begin{eqnarray}
\label{eq:bfieldsolution} \left(\begin{array}{c} \Psi_1(t_B)
\\ \Psi_{-1}(t_B)\end{array}\right) & = &
\left(\begin{array}{cc}
          \displaystyle{e^{i(\vartheta + K_B Z)}} & \displaystyle{0} \\
          \displaystyle{0} &
          \displaystyle{e^{-i(\vartheta + K_B Z)}}
\end{array}\right)
\left(\begin{array}{c} \Psi_1(0) \\ \Psi_{-1}(0)\end{array}\right)
\nonumber \\ & \equiv & M_B(\vartheta,K_B) \left(\begin{array}{c}
\Psi_1(0) \\ \Psi_{-1}(0)\end{array}\right),
\end{eqnarray}
where $\vartheta = -(\mu_B g_F/\hbar)B_0t_B$ and $K_B = -(\mu_B
g_F/\hbar)B't_B$ are the imprinted phase shift and phase gradient
(or wave vector), respectively. That is, the application of the
magnetic pulse results in a phase difference of $2\vartheta$
between the two Zeeman sublevels, and in addition it imparts them
wave vectors $\pm K_B$.

We note that although we have treated the pulsed excitations above
in an impulsive manner to illustrate their action, our simulations
describe the actions of the pulses correctly to the equations of
motion (\ref{eq:scaled}). The numerics confirm the accuracy of the
impulsive approximation for the parameters at consequence of the
fact that we consider pulse durations significantly shorter than
the soliton period (\ref{solper}) of $t_s=4.3$ ms.
\subsection{Gap soliton excitation}
\label{excitation}
\begin{figure}
\includegraphics*[width=0.95\columnwidth]{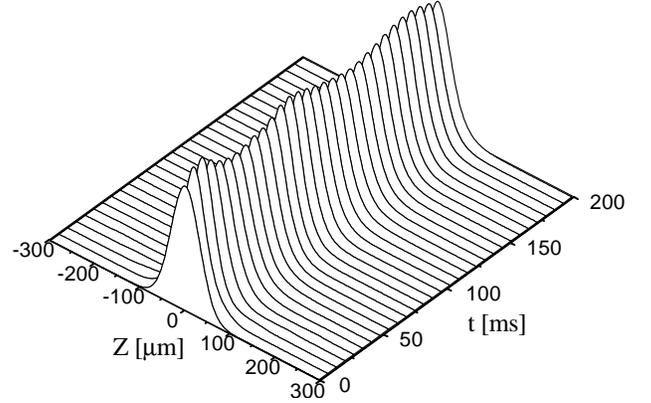}
\caption{The propagation of a soliton excited at $v=0$ over around
         50 soliton periods with initially 55000 atoms (shown is the
         total density). It displays
         oscillations in the peak density due to imperfect initial
         conditions. Further numerical investigation shows that these
         oscillations will eventually damp out after around $1 \mbox{
           s}$, accompanied by a loss of only two percent of the atoms
         from the soliton.}
\label{figexcitev00}
\end{figure}

To illustrate how stationary and moving solitons can be excited
using the proposed magneto-optical state scheme, we start from a
scalar condensate in the $|-1\rangle$ state,
$\Psi(Z,0)=[0,\Psi_0(z)]^T$, with spatial mode
\begin{equation}
\label{init} \Psi_0(Z)=\frac{N_s}{\sqrt{A_T}}\left (\frac{2}{\pi
W_s^2} \right )^{1/4} e^{-Z^2/W_s^2} ,
\end{equation}
with $W_s$ and $N_s$ the width and atom number of the gap soliton
desired, see Figs. \ref{figneta} and \ref{figwidtheta}.  
This Gaussian is chosen to
approximate the hyperbolic-secant structure of the analytic gap
soliton solution in Eq. (\ref{GapTot}).  For a stationary solution
we need to prepare the Zeeman sublevels with equal population and
with a $\pi$ phase difference. We further need to impose wave
vectors which are equal in magnitude but opposite in sign $K_{\pm
1}=\pm K_l$. This can be achieved by applying a laser pulse with
area $\chi=\pi/4$, followed by a magnetic pulse with
$\vartheta=\pi/4$ and $K_B=K_l$. The state then transforms as
$(t=t_p+t_B)$
\begin{eqnarray}
\label{eq:excitestationary} \Psi(Z,t) & = &
M_B(\pi/4,K_l)M_L(\pi/4)\Psi(Z,0) \nonumber \\ & = &
\frac{e^{\frac{3i\pi}{4}}} {\sqrt{2}}\left(\begin{array}{c}
e^{iK_l Z}\\ e^{-i\pi}e^{-i K_l Z}\end{array}\right)\Psi_0(Z).
\end{eqnarray}
Using the atomic parameters of Sodium, this situation can be
realized for a $10 \mbox{ }\mu\mbox{s}$ light pulse with
$I=2.69\mbox{ KW/cm}^2$, and a $200 \mbox{ }\mu\mbox{s}$ magnetic
field pulse with $B_0=0.89\mbox{ mG}$ and $B´=72.5\mbox{ G/cm}$
\cite{footnote1}. Fig. \ref{figexcitev00} shows the resulting
stable evolution of the total density $\left|\Psi(Z,t)\right|^2$.
As a result of the  Gaussian approximation to the exact solution
there are some slow oscillations imposed on the motion, but the
solution remains centered at $Z=0$ and stationary over a time
$t=200$ ms, much longer than the soliton period $t_s=4.3$ ms.

\begin{figure}
\includegraphics*[width=0.95\columnwidth]{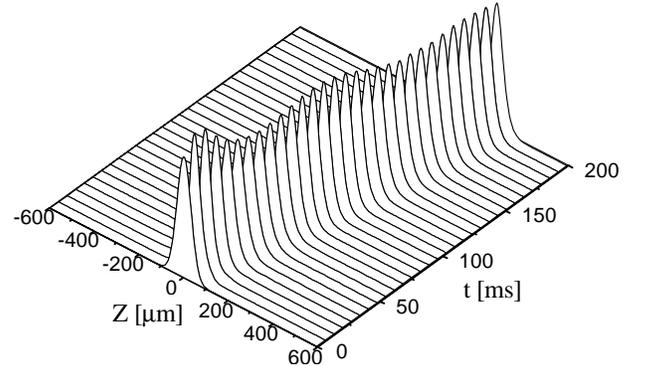}
\caption{The propagation of a moving soliton excited at a velocity of $0.18\mbox{ cm/s}$
         over around 50 soliton periods (shown is the total density).
         The actual velocity seen in
         the plot is very close to the excitation value. Peak density
         oscillation are again due to imperfect initial conditions.}
\label{figexcitev01}
\end{figure}

The excitation of a moving soliton is slightly more complicated
since the velocity dependent wave vector $K=-\gamma v/l_c$ in Eq.
(\ref{GapTot}) is no longer zero. To deal with this situation, we
first impart the wave vector $K$ to the initial $|-1\rangle$ state
in Eq. (\ref{init}) using a magnetic pulse with $\vartheta=0,
K_B=-K$, which for Sodium and a $200$ $\mu$s magnetic pulse can be
realized using $B_0=0, B^\prime=+0.094$ G/cm.  Here we
specifically take $v=0.1$, so that $K=-8306$ m$^{-1}$. We further
recall that moving solitons must have unequal populations of the
two Zeeman sublevels. For $v=0.1$, we have from  Eq. (\ref{Rpm})
that the ratio between the $|1\rangle$ and $|-1\rangle$
populations should be $1.22$. This is achieved by a coherent
optical coupling with $\chi=0.835$, and $I=2.69\mbox{ KW/cm}^2$,
corresponding to a pulse duration of $10.6$ $\mu$s. Finally, we
impart the wave vectors $\pm K_l$ and the $\pi$ phase difference
between the states the Zeeman sublevels with a magnetic pulse, as
for the stationary soliton above. Summarizing, the full
magneto-optical control sequence is described at $(t=2t_B+t_p)$ by
\begin{eqnarray}
\label{eq:excitemoving} \Psi(Z,t) & = &
M_B(\pi/4,K_l)M_L(\chi)M_B(0,-K)\Psi(Z,0) \nonumber \\ & = &
e^{\frac{3i\pi}{4}}\left(\begin{array}{c} \sin\chi e^{i(K+K_l
Z)}\\ \cos\chi e^{-i\pi}e^{i(K-K_l
Z)}\end{array}\right)\Psi_0(Z).
\end{eqnarray}
Fig. \ref{figexcitev01} shows the resulting numerical simulation
of a gap soliton with $v=0.1$ ($V_g=0.18$ cm/s), illustrating its
stable propagation is exhibited over many soliton periods.  We
have used the same scheme to launch gap solitons over the full
range of velocities.
\subsection{Soliton splitting}
\label{splitting}
The next application of magneto-optical control that we consider
is soliton splitting. To achieve this goal, we take advantage of
the fact that for fast solitons almost all of the population is in
one Zeeman sublevel, and the other state can be viewed as a small
perturbation. For example, for $v=0.5$ the ratio of the
populations is already $6$. This implies that the relative phase
between the two states is no longer of importance, so we need only
concentrate on getting the plane-wave factors right.

\begin{figure}
\includegraphics*[width=0.95\columnwidth]{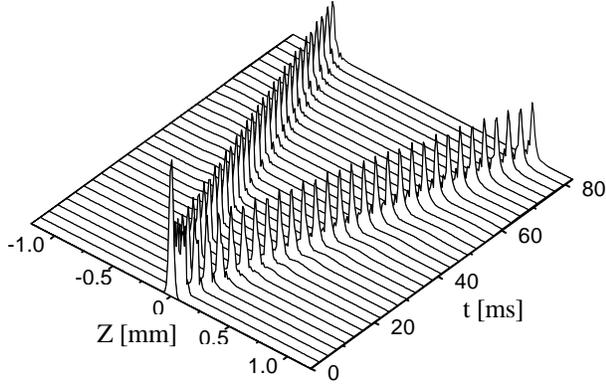}
\caption{The splitting of a condensate into two moving solitons,
         traveling into
         opposite directions (shown is the total density). Their
         velocity is approximately $\pm
         0.9\mbox{ cm/s}$, the propagation covers around 25 soliton periods.}
\label{figsplitting}
\end{figure}

\begin{figure}
\includegraphics*[width=0.95\columnwidth]{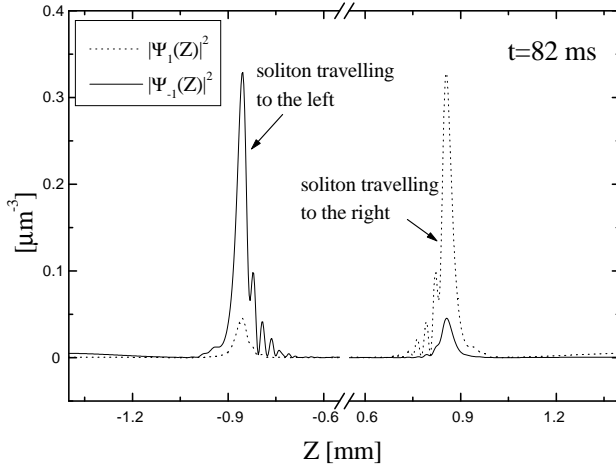}
\caption{Two solitons $82.0\mbox{ ms}$ after the splitting.
         The plot reveals that each soliton has still a small
         fraction of the other state bound to it.}
\label{figsplittingstates}
\end{figure}

Assume for concreteness that we start from an initial condition
with $N_0$ atoms in the $|-1\rangle$ state, and apply a laser
pulse of area $\pi/4$ to transfer half the population to the
$|+\rangle$ state. We can then apply a magnetic pulse to impose
the wave vectors $\pm (K_l+K)$, to the $|\pm 1\rangle$ states,
with $K=-\gamma v/l_c$ corresponding to a given velocity $|v|$.
Now if we choose $N_0=2N_s$, with $N_s$ the atom number for that
velocity, then for $|v|$ large enough we may expect to see
oppositely moving solitons emerge from the initial state. This is
illustrated in Fig. \ref{figsplitting}, which clearly illustrates
the emergence of two solitons with opposite velocities. Fig.
\ref{figsplittingstates}, which shows the density profiles for the
individual Zeeman sublevels for a time $t=82$ ms, confirms that
that each gap soliton indeed comprises a dominant Zeeman sublevel
plus a small component of the other state. During the early stages
of propagation, the solitons rearrange their phase and shape
before settling down. This is accompanied by some slowing down and
the familiar shedding of ``radiation.'' The emerging solitons are
therefore slower than their ``design velocity.'' For the specific
example of Fig. \ref{figsplitting}, the actual velocity is found
to be $1.08\mbox{ cm/s}$.
\subsection{Soliton reversal}
\label{reversal}
\begin{figure}
\includegraphics*[width=0.95\columnwidth]{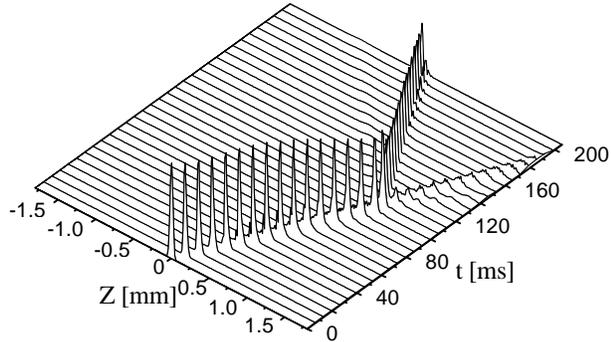}
\caption{The propagation of a fast soliton with velocity
         $v\approx1\mbox{ cm/s}$ whose direction is reversed
         after $100\mbox{ ms}$. The process of reversing causes
         small loss: a packet travels on to the right.}
\label{figreverse}
\end{figure}

In addition to offering the possibility of exciting moving gap
solitons, magneto-optical control can also reverse their direction
of propagation along $Z$. This can again be achieved by making use
of the fact that for fast solitons almost all the population is in
one Zeeman sublevel, so that we need only concentrate on getting
the plane-wave factors right. In the numerical simulation of Fig.
\ref{figreverse} we create a gap soliton with $v=0.59$ ($V_g=1$
cm/s) and let it propagate for $100$ ms. This soliton consists
essentially of a plane-wave factor $e^{i(K+K_l)z}$ multiplying
state $|1\rangle$. At $t=100$ ms we apply a laser pulse of area of
$\pi/2$ to transfer all the population to state $|-1\rangle$ and
change the phase to $e^{-i(K+K_l)z}$ using a magnetic field pulse.
This results in the same soliton as before the control sequence,
but with opposite velocity.  Note however the loss of some atomic
population to ``radiation'' in the process.
\section{Atomic Mach--Zehnder Interferometer}
\label{machzehnder}
As an illustration of the potential use of gap solitons employing
magneto-optical control here we consider a nonlinear atomic
Mach-Zehnder interferometer.  Solitons present some advantages for
atom interferometry in that they are many-atom wavepackets which
are immune to the effects of spreading, hence allowing longer path
lengths, and also reduced signal-to-noise for large atom numbers.
Typically many-body effects limit the utility of high-density
wavepackets due to spatially varying mean-field phase shifts, but
solitons have the cardinal virtue that they have fixed spatial
phase variations (for our situation this applies for faster
solitons). Thus solitons may provide a key to making maximal use
of high density sources for atom interferometry. Indeed they have
long been advocated for all-optical switching applications due to
these very properties.

\begin{figure}
\includegraphics*[width=0.95\columnwidth]{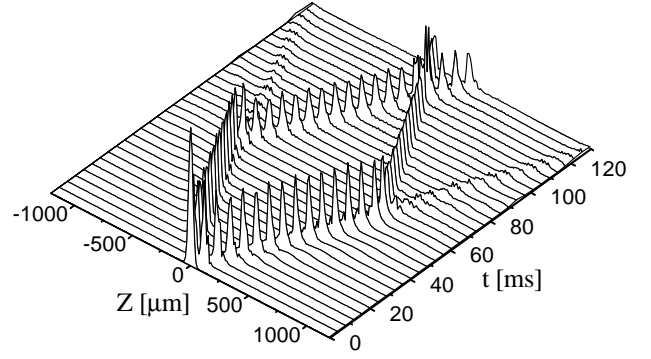}
\caption{Demonstration of an atomic Mach--Zehnder interferometer: The initial
         condensate is split into two counterpropagating fast
         solitons, then their direction is reversed and they collide
         (shown is the total density).}
\label{figmachzehnder}
\end{figure}

\begin{figure}
\includegraphics*[width=0.95\columnwidth]{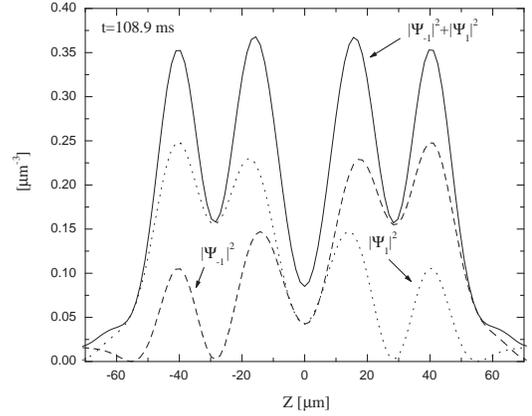}
\caption{Interference pattern when the two solitons collide in the
         Mach--Zehnder configuration: The
         total density (solid line) is symmetric, whereas the
         interference pattern of each of the two orthogonal Zeeman
         states (dotted and dashed lines) is
         asymmetric due to the shape of the colliding wavepackets. The
         contrast in the total density pattern is around 30\%.}
\label{figinterference}
\end{figure}

Our specific demonstration of a nonlinear atomic interferometer
based on solitons involves an initial scalar condensate that is
split into two oppositely moving solitons along $Z$, see Fig.
\ref{figmachzehnder} for $t<60$ ms.  At $t=60$ ms laser and
magnetic pulses are applied which act to reverse the direction of
the two solitons. The process of reversing causes some loss of
atoms in both solitons as before.  The two reversed soliton
components come together again at $t=120$ ms.  Since the colliding
solitons are predominantly in opposite orthogonal Zeeman
sublevels, the interference pattern appearing during the collision
is due to the contamination of each soliton by the other state,
see e.g. Fig. \ref{figsplittingstates}).  Fig.
\ref{figinterference} shows the interference at the soliton
collision in the total density (solid line) and also the
individual Zeeman sublevels, with a fringe contrast of around
30\%. These results demonstrate the potential use of gap solitons
for realizing nonlinear atom interferometers with high brightness
sources.
\section{Summary and outlook}
\label{conclusion}
Employing a spinor rather than a scalar condensate gives the
opportunity to externally manipulate bright atomic solitons by
conceptually simple magneto-optical methods without losing the
stable solitonic behavior. Specifically, the characteristic
properties of the gap soliton solutions lead to realistic
manipulation schemes that have been demonstrated explicitly in a
one-dimensional situation. This scheme has proven successful in
exciting solitons with different velocities at a minimal atom loss
rate as well as splitting, and reversing their direction of
propagation. The combination of these techniques resulted in the
demonstration of an atomic Mach--Zehnder interferometer, which
might be of interest in atom optical sensors. We remark that
although this was not explicitly discussed here, we have also
found numerical evidence for stable three--dimensional gap
solitons, when adding an optical potential for confinement in the
transverse direction.

It is now becoming amply evident that much of the future of atom
optics lies in integrated systems, or "atom optics on a chip" 
\cite{Dekk00}.
Optical and magnetic waveguides and beam splitters have recently
been demonstrated, but the coupling of a condensate into these
guides remains a major experimental challenge. The use of bright
solitons, with their potential to transport bright matter waves in
a controlled fashion, might offer a solution to this problem, and
merits further investigations. Further topics of interest include
the manipulation and control of bright solitons using phase
imprinting methods that employ the AC--Stark effect, as have
already been realized for dark
solitons.\cite{Dobr99,Dens00,Burg99}
\acknowledgements
\label{acknowledgements}
We have benefited from numerous discussions with M.\ G.\ Moore and
P.\ S.\ Jessen. The authors would also like to thank J.\ V.\
Moloney for CPU time. This work is supported in part by the U.S.\
Office of Naval Research under Contracts No.\ 14-91-J1205 and
No.\ 14-99-1-0806, by the National Science Foundation under Grant
No.\ PHY98-01099, by the U.S.\ Army Research Office, and by the
Joint Services Optics Program. S.P. was partly supported by the
German National Merit Foundation.
\appendix
\section{Four--Photon Coupling}
\label{appendix}
As mentioned in the Introduction, the Raman coupling between the
$| F_g = 1, m_g=-1\rangle$ and $| F_g = 1, m_g=1\rangle$ Zeeman
sublevels vanishes in the case of far-off resonance light, a
result of destructive interference with other $3P_{3/2}$ hyperfine
excited states \cite{Deut98}. In this limit the system reduces to
an effective $\left|J_g = 1/2\right\rangle \leftrightarrow
\left|J_e = 3/2\right\rangle $ transition where no $\Delta m =
\pm2$ transitions are allowed. This difficulty can be overcome by
adding a $\pi$--polarized laser beam from the side (e.g. along the
$X$-axis), which will not transfer any momentum along the
transverse direction. A proper choice of detunings allows one to
use another hyperfine ground state to mediate the $\Delta m =
\pm2$ transition, without populating it significantly.

\begin{figure}
\includegraphics*[width=0.95\columnwidth]{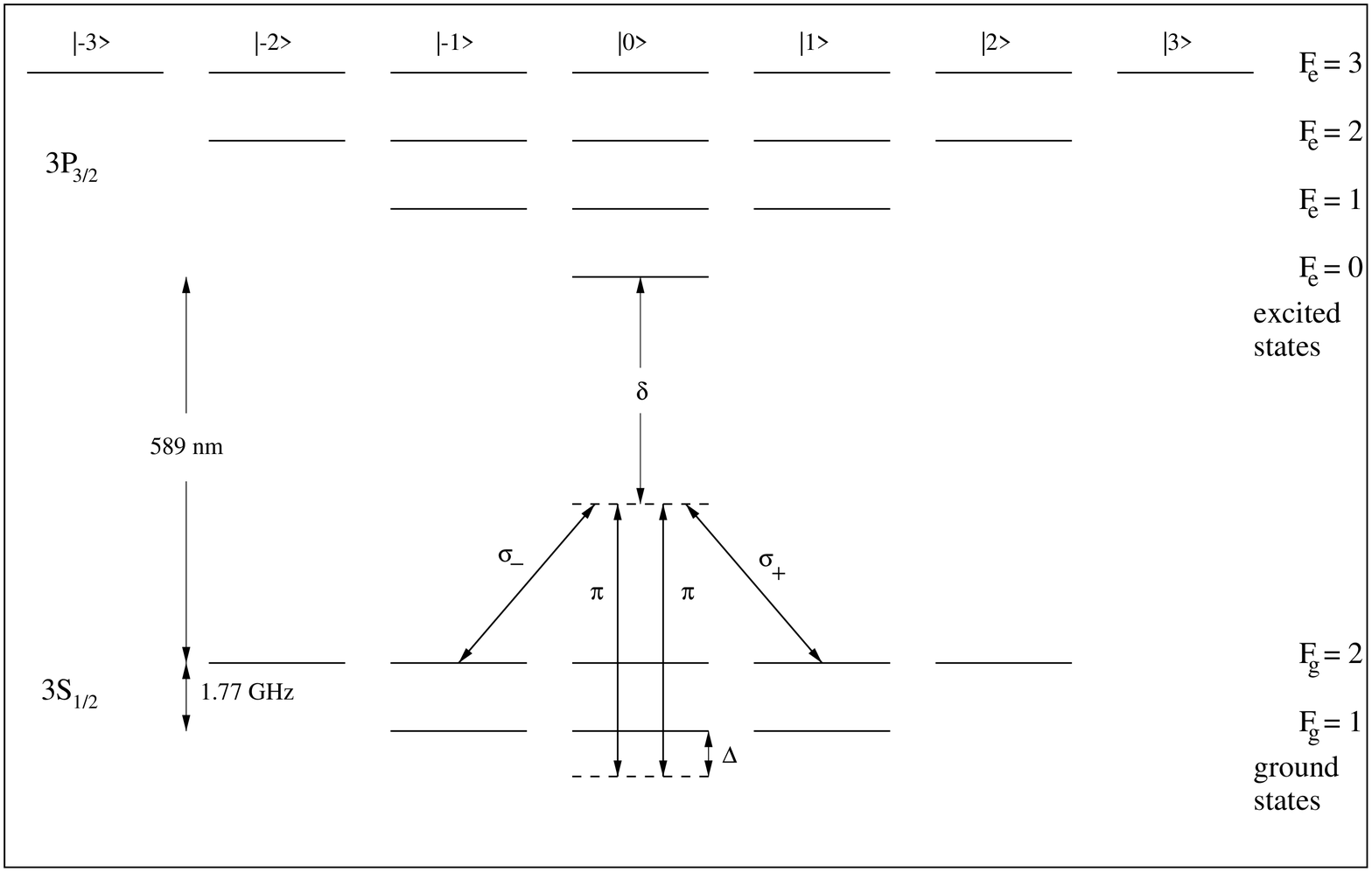}
\caption{Four--photon configuration: The $\left| F_g = 1, m = 0\right\rangle$ level mediates
  the coupling between the $\left| F_g = 2, m = \pm 1\right\rangle$ levels. The red
  detuning of the $\pi$--light ensures that all other levels are
  off--resonance. Although the state $\left| F_g = 1, m = 0
  \right\rangle$ is slightly populated, this ensures that we are
  sensitive to the hyperfine structure.}
\label{fig4photon}
\end{figure}

One disadvantage of this approach is that since this is now a
four-photon process, the intensity of the lasers must be increased
while still avoiding spontaneous emission. Choosing a frequency
difference between the $\pi$--light and the $\sigma$-light a few
$\mbox{MHz}$ larger than the splitting between the hyperfine
ground states as shown in Fig. \ref{fig4photon}, three time scales
govern the dynamics of the system. They are determined by (a) the
coupling of the ground state to the excited states, with detunings
in the $\mbox{THz}$ range; (b) the effective coupling between the
hyperfine ground states, with detunings in the $\mbox{GHz}$ resp.
$\mbox{MHz}$ range; (c) the desired effective $\Delta m = \pm2$
coupling. As it turns out, coupling the $\left| F_g = 2,
m_g=-1\right\rangle$ and $\left| F_g = 2, m_g=1\right\rangle$
levels is a better choice than staying in the $F_g = 1$ manifold,
as this minimizes the coupling to the outer states $\left| F_g =
2, m_g=\pm2\right\rangle$. Employing the intermediate $\left| F_g
= 1, m_g=0\right\rangle$ state makes the system aware of the
hyperfine structure and leads to a non--vanishing coupling.

Separating the time scales leads to an effective
coherent evolution between the $\left| F_g = 2,
m_g=\pm1\right\rangle$ states. It takes the same form as in Eq.
(\ref{eq:heff}), but with
\begin{eqnarray}
\label{eq:coeffeff}
g & \rightarrow & g_{\mathit{eff}}  = 1/192,\nonumber\\
\delta' & \rightarrow & \delta'_{\mathit{eff}} =
\frac{{\mathcal{D}}^4{\mathcal{E}}^2_\sigma{\mathcal{E}}^2_\pi}{\hbar^4\delta^2\Delta},
\end{eqnarray}
where ${\mathcal{E}}_\sigma$ is the amplitude for the $\sigma$--light
resp. ${\mathcal{E}}_\pi$ for the $\pi$--light and the detunings
$\delta$ and $\Delta$ are defined in Fig. \ref{fig4photon}.

The value of $g_{\mathit{eff}}$ involves Clebsch--Gordan and
Wigner--6J coefficients for all hyperfine transitions and
$\delta'_{\mathit{eff}}$ is clearly a four--photon term due to the
product of four electric field envelopes. Choosing the wavelength of
the $\sigma$--light 
$\lambda_{\delta} = 985 \mbox{ nm}$ as before,
$\Delta = \omega_{\pi}-\omega_{\sigma} = 1.85\mbox{ GHz}$ 
and laser intensities $I_{\sigma} = 100\mbox{ KW/cm}^2$
and $I_{\pi} = 158.3\mbox{ KW/cm}^2$ we obtain an effective coupling
$1/(g_{\mathit{eff}} \delta'_{\mathit{eff}}) = 685.0 \mbox{
}\mu\mbox{s}$, as in Sec. \ref{model}.

Numerical simulations of the full level scheme of Sodium is in
excellent agreement with the derived coupling in Eq.
(\ref{eq:coeffeff}) and shows that the intermediate level $| F_g =
1, m_g=0\rangle$ only contains around 0.1 percent of the atomic
population. All other levels are negligibly populated, they are
indeed far--off resonance. Since $| F_g = 1, m_g=0\rangle$ is a
ground state, spontaneous emission is not an issue.

The intensities might be reduced if $\delta$ were chosen smaller,
but that would of course alter the properties of the soliton,
since they depend on the wavelength of the optical fields. Note
that we cannot apply permanent magnetic fields to remove the
degeneracy of the hyperfine ground states and thus push
non--participating states out of resonance since this would
destroy the solitons. The light shifts induced by the lasers are
the same for each manifold and will not remove this degeneracy.

\begin{references}

\bibitem{Lenz93}
G.\ Lenz, P.\ Meystre, and E.\ W.\ Wright, 1993, Phys. Rev. Lett., {\bf 71},
3271--3274.

\bibitem{Deng99}
L.\ Deng, E.\ W.\ Hagley, J.\ Wen, M.\ Trippenbach, Y.\ Band,
P.\ S.\ Julienne, J.\ E.\ Simsarian, K.\ Helmerson, S.\ L.\ Rolston, and
W.\ D.\ Phillips, 1999, Nature, {\bf 398}, 218--220.

\bibitem{Inou99}
S.\ Inouye, A.\ P.\ Chikkatur, D.\ M.\ Stamper--Kurn, J.\ Stenger, D.\
E.\ Pritchard, and W.\ Ketterle, 1999, Science, {\bf 285}, 571--574.

\bibitem{Moor99a}
M.\ G.\ Moore and P.\ Meystre, 1999, Phys. Rev. Lett., {\bf 83}, 5202--5205.

\bibitem{Kozu99}
M.\ Kozuma, Y.\ Suzuki, Y.\ Torii, T.\ Sugiura, T.\ Kuga, E.\ W.\
Hagley, and L.\ Deng, 1999, Science, {\bf 286}, 2309--2312.

\bibitem{Scot73}
A.\ C.\ Scott, F.\ Y.\ F.\ Chu, and D.\ W.\ McLaughlin, 1973,
Soliton -- New Concept in Applied Science. {\em Proceedings of the IEEE}, 
{\bf 61}, 1443--1483.

\bibitem{Rein97}
W.\ P.\ Reinhardt and C.\ W.\ Clark, 1997, J. Phys. B, {\bf 30}, 
L785--L789.

\bibitem{Scot98}
T.\ F.\ Scott, R.\ J.\ Ballagh, and K.\ Burnett, 1998,
J. Phys. B, {\bf 31}, L329--L335.

\bibitem{Dum98}
R.\ Dum, J.\ I.\ Cirac, M.\ Lewenstein, and P.\ Zoller,
1998, Phys. Rev. Lett., {\bf 80}, 2972--2975.

\bibitem{Dobr99}
{\L}.\ Dobrek, M.\ Gajda, M.\ Lewenstein, K.\ Sengstock,
G.\ Birkl, and W.\ Ertmer, 1999, Phys. Rev. A, {\bf 60}, 
R3381--R3384.

\bibitem{Dens00}
J.\ Denschlag, J.\ E.\ Simsarian, D.\ L.\ Feder, C.\ W.\ Clark,
L.\ A.\ Collins, J.\ Cubizolles, L.\ Deng, E.\ W.\ Hagley, K.\
Helmerson, W.\ P.\ Reinhardt,
S.\ L.\ Rolston, B.\ I.\ Schneider, and W.\ D.\ Phillips, 2000,
Science, {\bf 287}, 97--101.

\bibitem{Burg99}
S.\ Burger, K.\ Bongs, S.\ Dettmer, W.\ Ertmer, K.\ Sengstock,
A.\ Sanpera, G.\ V.\ Shlyapnikov, and M.\ Lewenstein, 1999,
Phys. Rev. Lett., {\bf 83}, 5198--5201.

\bibitem{Chri89}
D.\ N.\ Christodoulides, and R.\ I.\ Joseph, 1989, Phys. Rev. Lett., 
{\bf 62}, 1746--1749.

\bibitem{Ster94}
C.\ M.\ de Sterke and J.\ E.\ Sipe, 1994, Gap Solitons.
In {\em Progress in Optics}, Vol. XXXIII, edited by E.\ Wolf 
(Amsterdam: Elsevier), pp. 203--260.

\bibitem{Zoba99}
O.\ Zobay, S.\ P\"otting, P.\ Meystre, and E.\ M.\ Wright, Phys. Rev. A,
1999, {\bf 59}, 643--648.

\bibitem{Key00}
M.\ Key, I.\ G.\ Hughes, W.\ Rooijakkers, B.\ E.\ Sauer, E.\ A.\
Hinds, D.\ J.\ Richardson, and P.\ G.\ Kazansky, 
Phys. Rev. Lett., 2000, {\bf 84}, 1371--1373.

\bibitem{Dekk00}
N.\ H.\ Dekker, C.\ S.\ Lee, V.\ Lorent, J.\ H.\ Thywissen,
S.\ P.\ Smith, M. Drndi{\'c}, R.\ M.\ Westervelt, and M.\ Prentiss.
Phys. Rev. Lett., 2000, {\bf 84}, 1124--1127.   

\bibitem{Zhan94}
W.\ Zhang, D.\ F.\ Walls, and B.\ C.\ Sanders, Phys. Rev. Lett., 1994, 
{\bf 72}, 60--63.

\bibitem{Deut98}
I.\ H.\ Deutsch and P.\ S.\ Jessen, Phys. Rev. A, 1998, {\bf 57}, 
1972--1986.

\bibitem{Hagl99a}
E.\ W.\ Hagley, L.\ Deng, M.\ Kozuma, J.\ Wen, K.\ Helmerson, S.\ L.\ Rolston,
and W.\ D.\ Phillips, 1999, Nature, {\bf 283}, 1706--1709.

\bibitem{Cohe92}
C.\ Cohen--Tannoudji, 1992, Atomic Motion in Laser Light. 
In {\em Fundamental Systems in Quantum Optics},
edited by J.\ Dalibard, J.\ M.\ Raimond, and J.\ Zinn--Justin
(Amsterdam: North--Holland), pp. 1--164.

\bibitem{Acev89}
A.\ B.\ Aceves and S.\ Wabnitz, Phys. Lett. A, 1989, {\bf 141}, 37--42.

\bibitem{Stam98}
D.\ M.\ Stamper--Kurn, M.\ R.\ Andrews, A.\ P.\ Chikkatur, S.\ Inouye,
H.--J.\ Miesner, J.\ Stenger, and W.\ Ketterle, Phys. Rev. Lett., 1998,
{\bf 80}, 2027--2030.




\bibitem{Moor99b}
M.\ G.\ Moore, O.\ Zobay, and P.\ Meystre,
Phys. Rev. A, 1999, {\bf 60}, 1491--1506.

\bibitem{Lenz94}
G.\ Lenz, P.\ Meystre, and E.\ W.\ Wright, Phys. Rev. A, 1994,
 {\bf 50}, 1681--1691.

\bibitem{footnote1}
Note that $g_F \approx -0.5$ for the $F_g = 1$ ground state of
Sodium.

\end{references}
\end{document}